\input{aipcheck}

\documentclass[
    ,final            
  ]
  {aipproc}

\layoutstyle{8x11double}
\newcommand\etal{{\it et~al.~}}
\newcommand\apj{ApJ}
\newcommand\apjl{ApJL}
\newcommand\apjs{ApJS}
\newcommand\aap{A\&A}
\newcommand\mnras{MNRAS}

\begin{document}

\title{Photoionization of Clustered Halos by the First Stars}


\author{Daniel Whalen}{
  address={Applied Physics (X-2), Los Alamos National Laboratory, Los Alamos, NM  87545, U.S.A.}
}

\author{Brian W. O'Shea}{
  address={Theoretical Astrophysics (T-6), Los Alamos National Laboratory, Los Alamos, NM  87545, U.S.A.}
}

\author{Joseph Smidt}{
  address={Department of Physics and Astronomy, Brigham Young University, Provo, UT  84602, U.S.A.}
}

\author{Michael L. Norman}{
  address={Center for Astrophysics and Space Sciences, University of California at San Diego, La Jolla, 
CA 92093, U.S.A.}
}


\begin{abstract}

We present numerical simulations of the photoevaporation of cosmological halos clustered around
a 120 M$_\odot$ primordial star, confining our study to structures capable of hosting Population 
III star formation.  The calculations include self-consistent multifrequency conservative transfer 
of UV photons together with nine-species primordial chemistry and all relevant radiative processes.  
The ultimate fates of these halos varies with central density and proximity to the central source 
but generally fall into one of four categories.  Diffuse halos with central densities below 2 - 3 
cm$^{-3}$ are completely ionized and evaporated by the central star anywhere in the cluster.  More 
evolved halo cores at densities above 2000 cm$^{-3}$ are impervious to both ionizing and Lyman-Werner
flux at most distances from the star and collapse of their cores proceeds without delay.  Radiative 
feedback in halos of intermediate density can be either positive or negative, depending on how the 
I-front remnant shock both compresses and deforms the core and enriches it with H$_2$.  We find that 
the 120 M$_\odot$ star photodissociates H$_2$ in most halos within the cluster but that catalysis by 
H- rapidly restores molecular hydrogen within a few hundred Kyr after the death of the star, with 
little delay in star formation.  Our models exhibit significant departures from previous one-dimensional 
spherically-symmetric simulations, which are prone to serious errors due to unphysical geometric 
focusing effects.

\end{abstract}

\maketitle


\section{Introduction}

Radiative feedback from one generation of stars on the next regulated the rise of stellar populations 
on many spatial scales since the highest redshifts. In early times this feedback came in two basic forms: 
ionizing UV photons and Lyman-Werner (LW) radiation between 11.18 eV and 13.6 eV that photodissociates 
molecular hydrogen.  Very massive primordial stars photoionized other cosmological halos clustered in 
their vicinity, in many instances preventing their collapse.  It is also believed that the H II regions 
of the first stars and protogalaxies established an ``entropy floor'' with elevated IGM temperatures that 
discouraged new star formation in general \citep{oh03}.  Prodigous sources of ionizing UV also emitted copious numbers 
of LW photons capable of sterilizing nearby halos of H$_2$, the key coolant permitting primordial gas to condense 
and form the first stars.  These photons propagated to much greater distances than ionizing UV because the 
IGM was transparent to them.  Over time this photodissociating background grew, permeating the early cosmos 
and suppressing H$_2$ formation in the first pregalactic objects.  These are three examples of 
negative radiative feedback.  On the other hand, molecular hydrogen can form in the outer layers of 
primordial I-fronts that later enhances cooling in the IGM \citep{rs01}, and such fronts can drive shocks into 
nearby cosmological halos, perturbing their cores and accelerating their collapse into stars.  How these 
two types of feedback competed to either promote or suppress subsequent star formation remains unclear.

\subsection{Numerical Models of Radiative Feedback}

Computer simulations of radiative feedback on star formation fall into three categories.  Those performed
in cosmological simulation volumes (1 Mpc$^3$ or more) follow large-scale structure formation without
radiative transfer, treating both ionizing and LW photons as a uniform background.  These 
studies suggest that metagalactic dissociating fluxes postpone but do not prevent primordial cloud collapse
into stars \citep{met01,met03,oshea07b,wa07}.  The second type are models that follow star formation on 
small scales in cosmological halos proximate to the first stars \citep{oet05}.  These surveys also sacrifice 
radiation transport to follow the detailed collapse of 
the halo into a new star.  Radiation is observed to exert both positive and negative feedback on primordial
cloud collapse in these studies.  More ambitious campaigns attempt to assemble the first primeval galaxies by 
the consecutive formation of primordial stars, one often in the relic H II region of its predecessor 
\citep{jet06,yet06}.  In these calculations, which emphasize radiative feedback in protogalaxy evolution, star 
formation in remnant ionized fields is weakly suppressed and, due to HD cooling, leads to a less massive 
generation of stars.

Simulations in the third and most recent category focus on the dissociation and evaporation of the halo rather
than its collapse into a star, inferring the likelihood of star formation from its final state 
\citep{su06,s07,awb07,as07}.  These
calculations find a variety of outcomes to halo photoevaporation, ranging from flash ionization and complete
eviction of baryons from diffuse halos to insulation from any radiation effects whatsoever in dense cores 
(collapse of the halo proceeds unimpeded in such cases).  Star formation in cosmological halos of intermediate 
densities can be either pre-empted or accelerated.  The most comprehensive survey to date was by \citet{as07}, 
who found local radiative feedback to be mostly neutral.  Unfortunately, this well-parametrized study is prone 
to unphysical geometrical focusing in its 1D Lagrangian grid, which leads to serious departures from the true 
evolution of the halos.

\section{Code Algorithm/Model Parameters}

We performed a suite of two-dimensional axisymmetric radiation hydrodynamical calculations of the ionization 
and photodissociation of a cosmological minihalo by a 120 M$_\odot$ star \citep{wet07}.  The halo is taken from 
an Enzo AMR code\footnote{\url{http://lca.ucsd.edu/portal/software/enzo}} \citep{enzomethod} simulation and spherically-averaged. 
The halo is then imported into the ZEUS-MP code\footnote{\url{http://lca.ucsd.edu/portal/software/zeus-mp2}} and 
photoevaporated.  ZEUS-MP \citep{wn06, wn07a} is a massively-parallel Eulerian reactive flow hydrocode with 
self-consistent multifrequency photon-conserving UV radiative transfer \citep{wn07b} coupled to nine-species 
primordial gas chemistry \citep{anet97}.  

\begin{figure}
  \includegraphics[height=.3\textheight]{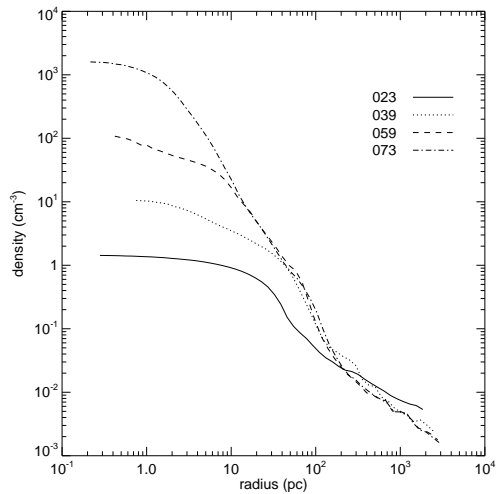}
  \caption{The four halo profiles.  The redshifts of the 023, 039, 059, and 073 profiles are 23.9, 
17.7, 15.6, and 15.0, respectively. \label{fig:halop}}
\end{figure}

The four evolutionary stages of the 1.35 $\times$ 10$^5$ M$_\odot$ primordial halo considered in our study 
are shown in Fig \ref{fig:halop}.  This halo mass was selected because it is the smallest in which a star 
would be expected to form.  Since the ionization front would have less impact on more massive halos, 
feedback effects would be most pronounced in this one.  Its central densities range from 1.43 cm$^{-3}$ to 
1596 cm$^{-3}$ from $z =$ 23.9 to 15.0.  We consider consecutive profiles from a single halo rather than 
sampling the entire cluster at a fixed redshift for three reasons.  First, the halos in the cluster have 
similar profiles and are largely coeval, so the emergent I-front may encounter a narrower range of density 
gradients at a single redshift than in one halo over a range of redshifts.  Second, the time at which the 
cluster is engulfed by the expanding H II region is an open parameter. Finally, we avoid constraining our 
results to a single cluster of halos and its associated properties (number of peaks, diameter, etc).  The 
range of radiative feedback in the entire cluster is better explored by evaporating a single halo whose 
central densities vary from low values that are easily ionized to high values in which core collapse would 
proceed uninterrupted.  Each profile was illuminated by a plane wave of photons from a 120 M$_\odot$ star 
at four distances: 150, 250, 500, and 1000 pc.  The plane wave was geometrically attenuated by 1/$R^2$, where
$R$ is the distance of the I-front from the star. The cloud is irradiated for 2.5 
Myr, the main sequence lifetime of the 120 M$_\odot$ star, and then left to evolve in the relic H II region 
an additional 2.5 Myr.  As a control we chemothermally evolved the halo for 4.8 Myr without radiation to 
determine the central densities to which the core condenses in the absence of stellar feedback.  Core 
densities in the 023 and 039 halos change by $\sim$ 1\%, those in the 059 halo rise 
from 108 cm$^{-3}$ to 200 cm$^{-3}$, and those in the 073 halo increase from 1600 cm$^{-3}$ to 6000 cm$^{-3}$.


\begin{figure}
  \includegraphics[height=.45\textheight]{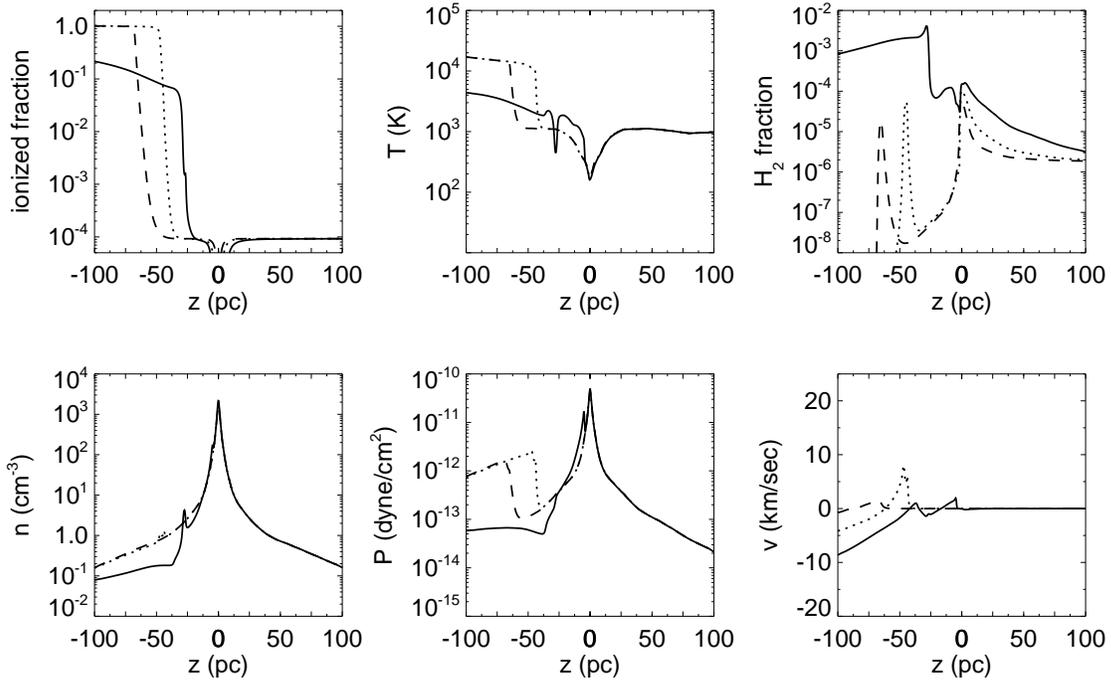}
  \caption{Ionized fraction, density, temperature, pressure, H$_2$ fraction and velocity profiles 
through the center of the 073 halo 500 pc from the star. Dashed line: 200 kyr (the R-type front), 
dotted line: 800 kyr (the D-type front), solid line: 5.0 Myr (the relic H II region). 
\label{fig:073500pcpr}}
\end{figure}

%

\begin{figure}
  \includegraphics[height=.4\textheight]{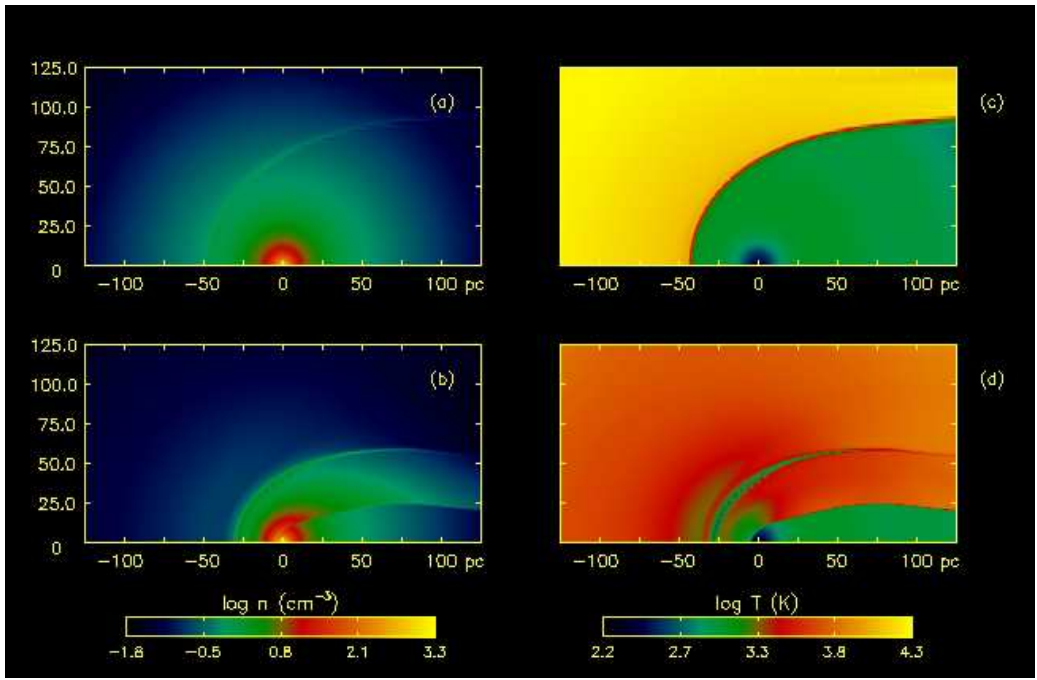}
  \caption{Panels (a) and (b) are density images of the 073 halo at 800 kyr and 5.0 Myr, respectively.  
Panels (c) and (d) are temperatures at 800 kyr and 5.0 Myr, respectively.\label{fig:073500pcrhoT}}
\end{figure}

%
\section{Fiducial Case}

We centered the 073 halo at the origin of a two-dimensional axisymmetric cylindrical coordinate box, 
with boundaries of -125 pc and 125 pc in $z$ and 0.01 pc and 125 pc in $r$.  The grid was discretized 
into 1000 zones in $z$ and 500 zones in $r$ for a spatial resolution of 0.25 pc.  Outflow conditions 
were applied to the upper and lower $z$ boundaries and reflecting and outflow conditions were assigned 
to the inner and outer boundaries in $r$, respectively.  The gas was primordial, 76\% H and 24\% He by 
mass. The gravitational potential of the dark matter was included by computing the potential necessary to 
cancel pressure forces everywhere on the grid (setting the halo in hydrostatic equilibrium) and holding
this potential fixed throughout the simulation.  The self gravity of the gas was computed every 
hydrodynamical time step with a two dimensional conjugate gradient Poisson solver.  Cooling by electron collisional excitation 
and ionization, recombination, bremsstrahlung, and inverse Compton scattering of the cosmic microwave 
background (CMB), assuming a redshift $z = $ 20, was present in all the models.  H$_2$ cooling was also 
included using the cooling curves of \citet{gp98}. For simplicity, we assume ionization and H$_2$ fractions 
of 1.0 $\times$ 10$^{-4}$ and 2 $\times$ 10$^{-6}$, respectively, appropriate for the IGM at z $\sim$ 20.

\begin{figure}
  \includegraphics[height=.6\textheight]{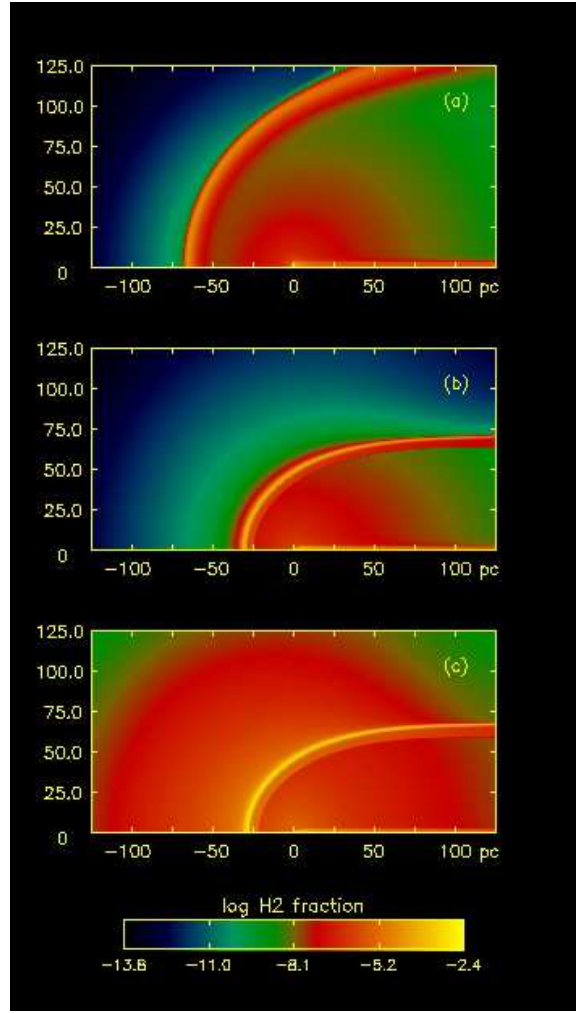}
  \caption{073\_500pc model: H$_2$ fractions.  Panels (a), (b) and (c) are at 210 kyr, 
2.29 Myr, and 2.59 Myr, respectively.\label{fig:073500pcH2}}
\end{figure}

In Fig \ref{fig:073500pcpr} we show profiles of ionized fraction, density, temperature, pressure, H$_2$ 
fraction, and velocity along the z-axis at three stages of photoevaporation of the 073 halo, with 
n$_c$ = 1596 cm$^{-3}$ and the star 500 pc from the center of the halo.  At 200 kyr the ionization front 
is R-type (becoming D-type 50 pc from the center of the cloud).  Lyman-Werner photons passing 
through the front partially dissociate the outer halo but the core itself remains deeply shielded 
and molecular hydrogen levels rise rapidly there.  Panel (a) of Fig \ref{fig:073500pcH2} shows that the 
core blocks LW photons in a $\sim$ 10 pc band centered on the z-axis.  Central H$_2$ fractions rise from 
2 $\times$ 10$^{-6}$ to 1 $\times$ 10$^{-4}$ before the front becomes D-type due to catalysis by free 
electrons, leveling off thereafter for 
the lifetime of the star.  The partially ionized warm gas in the outer layers of the front also 
catalyzes H$_2$ production, visible as the bright yellow parabolic arc in all three panels of Fig 
\ref{fig:073500pcH2}.  H$_2$ fractions in the R-type front climb to 2 $\times$ 10$^{-4}$ and remain 
steady after the front transforms to D-type.  

Density and temperature images of the D-type front at 800 kyr and relic H II region at 5 Myr are shown in
Fig \ref{fig:073500pcrhoT}.  The front has a cometary appearance because it preferentially advances in the
more tenuous outer radii of the halo.  Comparison of panels (a) and (b) reveals that the shadow is axially 
compressed by the ionized gas surrounding it but the cylindrical shock never reaches 
the axis in this run.  Perturbations are visible in the front above the axis at 800 kyr.  These fluctuations 
may be an early stage of instability that arises in D-type fronts when UV photons are oblique to the 
shock \citep{rjw02}.  The perturbations are longer further from the central axis where radiation is incident 
to the front at smaller angles.  LW photons preferentially stream through the underdensities, dissociating 
H$_2$ beyond in bands visible between z = 50 pc and z = 100 pc at 2.29 Myr in panel (b) of Fig 
\ref{fig:073500pcH2}.  These features are transient, vanishing by the time the star dies.  The shock 
decelerates from 10 km/sec at 50 pc to 7.5 km/sec 25 pc from the core at 2.5 Myr.  


When recombination commences at 2.5 Myr molecular hydrogen forms most rapidly in the shock but then 
radiates outward down the density gradient in the ionized regions of the halo, as shown at 2.59 Myr 
in Fig \ref{fig:073500pcH2}.  H$_2$ fractions rise more gradually in the core over the 
following 2.4 Myr, by approximately 50\%.  The 5.0 Myr curves in Fig \ref{fig:073500pcpr} show
that the center of the cloud is impervious to recombination flows: the incoming shock (eventually 
approaching to within a few pc), the rarefaction wave, ionized backflows at larger radii, and the 
compression of the shadow toward the axis.  One interesting feature of the relic H II 
region is the appearance of Rayleigh-Taylor instabilities at the interface of the warm ionized gas 
and shock remnant in panels (b) and (d) in Fig \ref{fig:073500pcrhoT}.  The fingers of partially 
ionized (and rapidly cooling) gas lengthen along the arc away from the axis, but do not affect
the dynamics of the core.

Cooling continues in the center of the halo throughout the simulation, with core temperatures falling 
from 200 K to 100 K and densities rising from 1596 cm$^{-3}$ to $\sim$ 3000 cm$^{-3}$ over 5 Myr.  The 
shock remnant eventually ripples through the center but with only minor density fluctuations.  We find 
that the core continues to contract while the front photoionizes the halo, reaching final densities 
similar to those when no star is present.  Star formation proceeds unimpeded in this halo.

\section{Results and Conclusions} 

\begin{figure}
  \includegraphics[height=.35\textheight]{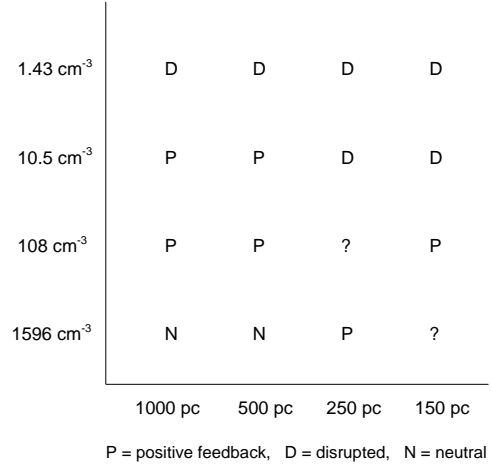}
  \caption{Radiative feedback on star formation in the halos.\label{SF}}
\end{figure}

Final outcomes for the other 15 models fell into four categories:

\begin{itemize}
\item
Relatively dense halo cores are shielded from ionizing and LW UV.  H$_2$ in these cores  
is protected from Lyman-Werner flux even without the H$_2$ formed in the front.  Collapse of the core 
into a star proceeds without delay.
\item
In the other extreme, diffuse halos are easily destroyed by the radiation of the star wherever they
reside in the cluster, either by an R-type front that ionizes the cloud on time scales much shorter
than its dynamical time or by a D-type front that snowplows gas from the core at speeds greater than
its escape velocity.
\item
Within a band of fluxes and central densities, the ionization front can jostle the core of the halo with a 
shock enriched with H$_2$, compressing and speeding its collapse without destroying it.  H$_2$ is initially
photodissociated in the halos but rapidly reconstitutes after the death of the star, with little delay in
cooling.
\item
In a less forgiving sector of the flux-n$_c$ plane the shock remnant displaces the core more violently 
and the implosion of the shadow deforms it more severely, but it still survives.  Their effect on star 
formation is uncertain.
\end{itemize}
We summarize these results in Fig \ref{SF}.

\begin{theacknowledgments}
DW thanks Tom Abel, Kyungjin Ahn, Greg Bryan and Alex Heger for helpful discussions concerning these 
simulations.  This work was carried out under the auspices of the National Nuclear Security 
Administration of the U.S. Department of Energy at Los Alamos National Laboratory under Contract No. 
DE-AC52-06NA25396.  The simulations were performed at SDSC and NCSA under NRAC allocation MCA98N020 
and at Los Alamos National Laboratory.
\end{theacknowledgments}

\bibliographystyle{aipproc}   

\bibliography{sample}

\IfFileExists{\jobname.bbl}{}
 {\typeout{}
  \typeout{******************************************}
  \typeout{** Please run "bibtex \jobname" to optain}
  \typeout{** the bibliography and then re-run LaTeX}
  \typeout{** twice to fix the references!}
  \typeout{******************************************}
  \typeout{}
 }


\end{document}